\documentclass[aps,prl,groupedaddress,twocolumn,10pt,reprint,superscriptaddress,nofootinbib,longbibliography,floatfix]{revtex4-2}
\usepackage{epsfig,dsfont,amssymb,amsmath,amsthm,amsfonts,amsbsy,mathrsfs} 
\usepackage{lipsum}
\usepackage{color}
\usepackage{graphicx}
\usepackage{amsmath}
\usepackage{amssymb}
\usepackage{subfigure}
\usepackage{float}
\usepackage{hyperref}
\usepackage{verbatim}
\usepackage{soul}
\usepackage{CJK}
\setstcolor{red}

 \usepackage{gensymb}
 \usepackage{comment}
\renewcommand{\eqref}[1]{\textrm{Eq.}(\ref{#1})}

\newcommand {\be}{\begin{equation}}
\newcommand {\ee}{\end{equation}}

\begin{document}

\title{Contact-Dependent Allosteric Ion Gating Shapes Directional Asymmetry in the Bacterial Flagellar Motor}

\author{Jiading Zhu}
\thanks{These authors contributed equally to this work.}
\affiliation{Department of Physics, Tsinghua University, Beijing, 100084, China }

\author{Yongnan Hu}
\thanks{These authors contributed equally to this work.}
\affiliation{Department of Physics, Tsinghua University, Beijing, 100084, China }

\author{Yuhai Tu}
\email{ytu@flatironinstitute.org}
\affiliation{Center for Computational Biology and Center for Computational Neuroscience, Flatiron Institute, New York,
New York 10010, USA}

\author{Yuansheng Cao}
\email{yscao@tsinghua.edu.cn}
\affiliation{Department of Physics, Tsinghua University, Beijing, 100084, China }

\begin{abstract} 

The bacterial flagellar motor (BFM) is a molecular machine that rotates either counterclockwise (CCW) or clockwise (CW), powered by the transmembrane ion electrochemical potential. A longstanding puzzle is the directional asymmetry of its torque–speed relation, which is concave in CCW but nearly linear in CW. Recent cryo-EM structures revealed a gear-like rotor–stator architecture in which rotational switching is driven by changes in the rotor–stator contact configuration. Inspired by these structural insights, we develop a stochastic mechanochemical model that couples rotor–stator mechanics to ion translocation kinetics and constrain it using physiological torque–speed measurements together with data on rotor–stator relative motion. Our analysis shows that, under physiological conditions, the motor operates in a tight-engagement regime in which the torque–speed relation is largely insensitive to the detailed form of rotor–stator mechanical interactions, arguing against direction-dependent mechanics as the origin of the observed asymmetry. We instead propose a contact-dependent gating mechanism in which MotA–FliG interactions allosterically regulate ion release from the MotB acidic site. Molecular dynamics simulations support this mechanism by revealing direction-dependent MotA–MotB acidic-site configurations, with the CCW state preferentially stabilizing an ion-release-competent geometry. Incorporating this gating mechanism quantitatively reproduces the observed asymmetry: stronger gating in the CCW state shortens torque-free waiting phases, enhances torque generation, and produces a concave torque–speed relation, whereas weaker gating in the CW state yields lower torque and an approximately linear torque–speed relation. These results establish a mechanistic link between structural asymmetry and motor function and provide testable predictions for mutagenesis and targeted perturbation experiments.



\end{abstract}
\maketitle


\section{Introduction}

The bacterial flagellar motor (BFM) is a complex molecular machine that powers the swimming of many motile bacteria. It functions as a rotary engine embedded in the bacterial cell wall and cytoplasmic membrane and is driven by an ion motive force (IMF; e.g., $H^+$ or $Na^+$), an electrochemical potential difference of ions across the membrane~\cite{berg1973,Larsen1974,Hirota1981,Berg2003}. In \textit{Escherichia coli}, the BFM rotates bidirectionally, enabling two distinct modes of bacterial movement~\cite{wadhwa:2021:nat_rev_micro}. During counterclockwise (CCW) rotation, multiple flagellar filaments form a coherent bundle that propels the bacterium forward in a motion known as a ``run''. In contrast, clockwise (CW) rotation disrupts the bundle, causing the bacterium to undergo a ``tumble'', which randomly reorients the cell. Alternating between runs and tumbles allows bacteria to navigate chemical gradients and move toward favorable environments or away from harmful ones, a process known as chemotaxis.

Torque in the BFM is generated through interactions between rotor and stator components~\cite{blair1988,blair1990mota,kojima2004s,Yorimitsu2004,Chun1988,wadhwa:2019:pnas,wadhwa:2021:nat_rev_micro}. The rotor consists of the membrane-embedded MS ring and a cytoplasmic C ring composed of multiple copies of the proteins FliG, FliM, and FliN. Each stator unit, embedded in the membrane, contains a MotB dimer surrounded by a ring of MotA subunits~\cite{Asai1997,Sato2000,Roujeinikova2008,Block1984,blair1988,Lee2016}. Recent cryo-EM studies revealed that the MotA ring forms a pentamer~\cite{deme2020,santiveri2020}. Each stator contains two ion channels associated with the two MotB subunits. Ion translocation (e.g., proton $H^+$ flow) through these channels drives the rotation of the MotA ring, which in turn exerts torque on the rotor through interactions with FliG.

The cogwheel-like architecture of the motor and structures captured in different rotational states have revealed an elegant mechanism for directional switching. The stator complexes rotate unidirectionally, while conformational changes in FliG reverse the direction of rotor motion. In particular, FliG interacts with the inner edge of the stator during CCW rotation but changes conformation to interact with the outer edge during CW rotation~\cite{chang2020,carroll2020} (Fig.~1A).

A key functional property of the BFM is its torque--speed relation. A longstanding puzzle is the asymmetry between CW and CCW rotation\cite{Xing2006,Mandadapu2015,tu2018design}. During CCW rotation, the motor exhibits a concave torque--speed curve: torque remains nearly constant over a broad range of speeds and drops only at high speeds~\cite{Chen2000,Lo2013}. This behavior enables the motor to maintain high torque across a wide range of loads during swimming. In contrast, the torque--speed relation for CW rotation is approximately linear, with torque decreasing roughly linearly with speed~\cite{Yuan2010}, reminiscent of linear molecular motors such as kinesin~\cite{svoboda1994force,Lipowsky2007}. Despite this difference in shape, both rotational states share the same stall torque and maximum speed~\cite{yuan2008,wang2022direct,Lo2013}. Although load-dependent stator remodeling can influence measured motor output, recent CW/CCW remodeling measurements\cite{wadhwa:2021:pnas,lele:2013:pnas} suggest that the concave CCW and linear CW torque--speed curves reflect direction-dependent single-stator mechanochemistry rather than a remodeling effect.

These observations thus raise a natural question: can recent structural insights into the BFM reveal the mechanistic origin of this CW--CCW asymmetry? In this work, we develop a mechanochemical model of the BFM~\cite{cao2022modeling} based on cryo-EM structures of the stator complex and the distinct conformational states of FliG associated with CW and CCW rotation~\cite{santiveri2020,deme2020,chang2020,tan2024,johnson2024,singh2024}. Using this structure-based model, we investigate how stator--rotor interactions and ion translocation dynamics shape the torque--speed relation. By combining theoretical analysis, numerical simulations, and molecular dynamics studies of the ion release pathway, we uncover a contact-dependent gating mechanism for ion release that explains the observed CW--CCW asymmetry.

\section{The gear-to-gear model of rotary bacterial flagellar motor}
\begin{figure}
    \centering    \includegraphics[width=1.1\linewidth]{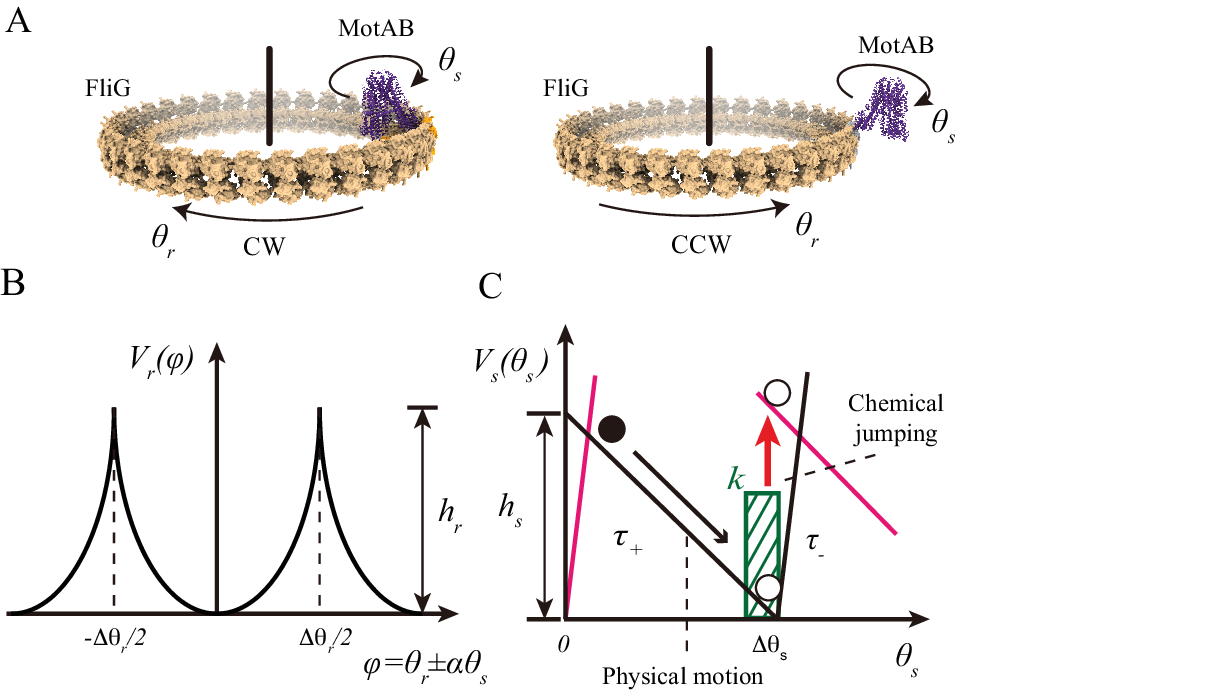}
    \caption{Model of two coupled nano-rings. (A) Rotor-stator coupling in CW (left) and CCW (right) rotations. In CCW rotation the stator (MotAB) engages the rotor's outer periphery, whereas in CW rotation it engages the rotor's inner periphery. (B) Periodic rotor-stator interaction potential $V_r$ with depth $h_r$. (C) Periodic stator free-energy landscape $V_s$. The stator generates positive torque $\tau_+=h_s/\Delta\theta_s$, and jumps to the next cycle with rate $k(\theta_s)$. Here we only show the ``no-gating" ion kinetics in the CW mode.}
    \label{fig1}
\end{figure}

We briefly describe the mechanochemical model of the bacterial flagellar motor (BFM), which consists of two coupled components: the mechanical motion of the motor and the chemical transitions driven by ion translocation.

First, rotations of the rotor and a stator can be described by two coupled overdamped Langevin equations:
\begin{equation}
    \xi_{r}\frac{d\theta_{r}}{dt}=-\frac{\partial}{\partial\theta_{r}}V_{r,\sigma}(\varphi_\sigma)+\sqrt{2\xi _{r}}\eta_1(t),
\end{equation}
\begin{equation}
    \xi_{s}\frac{d\theta_{s}}{dt}=-\frac{\partial}{\partial\theta_{s}}V_{r,\sigma}(\varphi_\sigma)-\frac{\partial}{\partial\theta_{s}}V_{s}( \theta_{s})+\sqrt{2\xi _{s}}\eta_2(t),
\end{equation}
where $\sigma=1$ and $-1$ denote the CCW and CW rotational states of the motor, respectively. $\theta_r$ and $\theta_s$ represent the angular coordinates of the rotor and stator. The relative angle between them is $\varphi_{\sigma}=\theta_r+\sigma\alpha\theta_s$ with $\alpha = 5/26$ the gear ratio. $V_{r,\sigma}$ is the stator-rotor interaction potential (Fig.~1B). 
We assume that $V_{r,\sigma}$ depends only on the relative coordinate $\varphi_{\sigma}$, which ensures periodicity in both $\theta_r$ and $\theta_s$. 
$V_s(\theta_s)$ is the interaction potential between MotA and MotB within the stator with period $\Delta\theta_s=2\pi/5$. For simplicity, we adopt a piece-wise linear form for $V_s$ with height $h_s$. The slope $\tau_\pm (\propto h_s)$ of $V_s(\theta_s)$ corresponds to the instantaneous torque generated by the stator on each side of the potential minimum   (Fig.~1C). $\xi_r$ and $\xi_{s}$ are viscous loads for the rotor and the stator, 
and $\eta_{1,2}(t)$ represent thermal noise satisfying $\langle\eta_i(t)\eta_j(t')\rangle=k_B T\delta_{ij}\delta(t-t')$ with $k_BT$ the thermal energy. 

Second, ion-driven chemical transitions provide the active component of the motor dynamics. As illustrated in Fig.~1C, an IMF-powered chemical transition (red arrow) moves the system to a higher stator potential energy level in a different energy landscape (red curve). These transitions are triggered by ion translocation events, which occur at a rate $k_{\sigma}(\theta_s)$ that depends on the relative configuration between MotB and the MotA ring:
\be
\text{Prob(transition during } t\to t+dt)
=k_\sigma(\theta_s)dt.
\ee
Backward transitions are neglected, which is justified under sufficiently large ion motive force (IMF).

Cryo-EM studies show that the stator interacts with the outer and inner periphery of the rotor in the CCW and CW states, respectively (Fig.~1A). The distinct contacts between FliG in the rotor and MotA in the stator can therefore lead to different rotor–stator interaction potentials $V_{r,\sigma}(\varphi_\sigma)$ as well as different ion-gated transition rates $k_{\sigma}(\theta_s)$.

In the remainder of the paper, we use this model together with experimental data to investigate the molecular mechanisms that may underlie the distinct torque–speed relations observed for the CCW and CW states of the motor. For notational simplicity, we drop the explicit index $\sigma$ in the following and focus on the general question of whether, and how, different forms of the rotor–stator interaction $V_r$ and the transition rate $k$ can produce different torque–speed dependencies.

\begin{figure}
    \centering
    \includegraphics[width=1.0\linewidth]{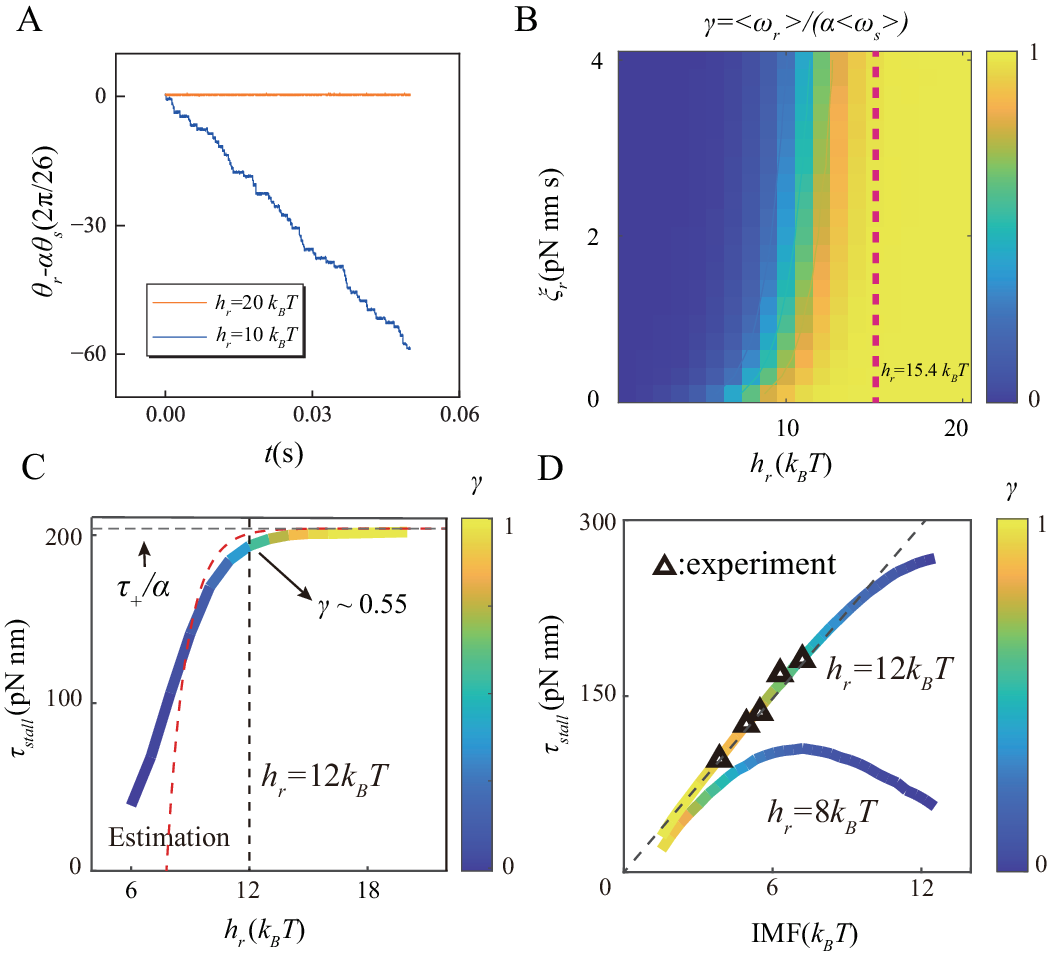}
    \caption{The stator-rotor slippage and the stall torque. (A) Trajectories of the relative angle $\varphi=\theta_r-\alpha\theta_s$ in the deterministic-slipping (blue) and stochastic-slipping (red) regimes. (B) Engagement fraction $\gamma=\left \langle \omega_{r} \right \rangle / \alpha\left \langle \omega_{s} \right \rangle$ as a function of rotor load $\xi_r$ and rotor-stator coupling depth $h_r$ (with $h_s=12k_BT$). Red dashed line: $h_r\approx 15.4 k_BT$ inferred from experiment\cite{hosu2025torque}. 
    (C) Stall torque versus \(h_{r}\) ($\xi_r\approx 3.3$ pN nm s for 1$\mu$m bead assay). Increasing \(h_{r}\) suppresses slipping and drives the stall toward the upper bound \(\tau_{+}/\alpha\). The torque reduction follows an exponential decay (dashed line, Eq.~\ref{slipping}). At $h_r=12k_BT$, $\gamma \sim 0.55$ while the torque drop is $\sim 3\%$. (D) Stall torque versus IMF. Assuming $\tau_+$ and $k$ scale linearly with IMF, $\tau_{\text{stall}}$ rises linearly at low IMF (dashed line) but bends downward upon entering the determinist-slipping regime. Experimental data (triangles, from \cite{Lo2013}) remain linear, implying tight engagement and $h_r\geq 12k_BT$.}
\end{figure}

\section{Rotor and stator are strongly coupled with minimal torque reduction at stall}

Because the BFM forms a two–cogwheel system, slippage between the active stator and the passive rotor can occur depending on the rotor–stator interaction strength ($h_r$) and the motor load $\xi_r$. As illustrated in Fig.~2A, the relative angle $\varphi=\theta_r-\alpha\theta_s$ fluctuates around zero when the interaction is strong ($h_r=20k_BT$), indicating tight engagement between rotor and stator. In contrast, for a weaker interaction ($h_r=10k_BT$), $\varphi$ decreases approximately linearly with time, indicating substantial stator–rotor slipping.

Slippage can be quantified by the engagement fraction $\gamma=\langle\omega_r\rangle/\alpha\langle\omega_s\rangle$, which measures the fraction of time that the rotor and stator remain tightly engaged. As shown in Fig.~2B, engagement weakens (smaller $\gamma$) as the load $\xi_r$ increases or the interaction strength $h_r$ decreases, with a relatively sharp transition from tight to loose coupling. The results in Fig.~2B were obtained using a quadratic $V_r$; other forms of $V_r$ produce qualitatively similar behavior (see SI and Fig.~S1 for details).

From Eqs.~1\&2, the dynamics of the relative coordinate $\varphi$ obey
\begin{equation}
\dot{\varphi} = -\frac{V_r'(\varphi)}{\xi} + \frac{\alpha V_s'(\theta_{s})}{\xi_{s}} + \sqrt{\frac{2}{\xi}}\eta(t),
\label{relative_angle}
\end{equation}
where $\eta(t)$ denotes thermal noise and the effective friction is $\xi = (\frac{1}{\xi_{r}}+\frac{\alpha^{2}}{\xi_{s}})^{-1}$. \eqref{relative_angle} shows that dynamics of $\varphi$ is governed by an effective potential $V_{eff}=V_r/\xi -\alpha V_s/\xi_s$, which reflects the competition between the rotor–stator interaction $V_r$, which tends to maintain engagement, and the stator potential $V_s$, which drives stator motion. 

In the strong stator-rotor interaction regime ($h_r > h_s/2$), a stable fixed point $\varphi^*$ exists at the minimum of $V_{eff}$ when $V_r'/\xi=\alpha V'_s/\xi_s$. 
Thermal fluctuations can nevertheless induce rare barrier-crossing events that transiently break engagement; we refer to these as stochastic slipping. In the high-load limit ($\xi_r\gg\xi_s$), the mean slipping rate can be approximated by Kramers’ escape rate:
\begin{equation}
    \langle\dot\varphi\rangle\approx -\frac{\alpha\tau_{+}}{\xi_{s}}e^{-\Delta h/k_BT},
    \label{slipping_rate}
\end{equation}
where $\Delta h\equiv \int_0^{\varphi^*}(V_r'-V_s')d\varphi\sim h_r-h_s/2$ is the energy barrier (see SI and Fig.~S2 for details).

When the interaction is weak ($h_r < h_s/2$), the barrier disappears ($\Delta h < 0$), and $\varphi$ decreases monotonically, as shown in Fig.~2A. We term this the deterministic slipping regime, in which the rotor persistently lags behind the stator.

In the stochastic-slipping regime ($e^{-\Delta h/k_BT}\ll1$), thermally induced disengagement events slightly reduce the stall torque $\tau_{\rm stall}$ from its theoretical maximum $\tau_{\rm max}\equiv\tau_+/\alpha$. Defining $\Delta\tau=\tau_{\rm max}-\tau_{\rm stall}$, the relative torque reduction in the tight-coupling regime can be approximated as (see SI for detailed derivation):
\begin{equation}
\frac{\Delta \tau}{\tau_{\text{max}}} \approx \left(1+\frac{t_w}{t_f}\right)e^{-\frac{\Delta h}{k_BT}},
\label{slipping}
\end{equation}
where $t_w=(\int_0^{\Delta\theta_s}k(\theta)P_{ss}(\theta)d\theta)^{-1}$ is the average waiting time for the chemical transition and $t_f=\xi_s\Delta\theta_s/\tau_+$ is the time for a free stator to traverse the torque-generating region $\Delta\theta_s$.

Simulation results (Fig.~2C) are consistent with \eqref{slipping}. In the strong-coupling regime ($h_r>h_s/2$), the stall torque approaches $\tau_{\rm max}$ and becomes largely insensitive to $h_r$. In contrast, in the weak-coupling regime ($h_r\ll h_s/2$), $\tau_{\rm stall}$ decreases sharply with $h_r$ and vanishes as $h_r\to0$. Our analysis also yields an analytical expression for the engagement fraction in the strong-coupling limit (SI):
\begin{equation}
\gamma \approx \frac{1-e^{-\Delta h/k_{B}T}}{1+\frac{\alpha^{2}\xi_{r}}{\xi_{s}}e^{-\Delta h/k_{B}T}}.
\label{gamma-deltah}
\end{equation}

We next determine whether the BFM operates in the stochastic- or deterministic-slipping regime by combining our model with experimental data. Experiments with varying IMF by Lo et al.~\cite{Lo2013} suggest that both $h_s$ (thus $\tau_+$) and the transition rate $k$ scale approximately linearly with IMF, which  we adopt in our model (SI, Fig.~S3). As shown in Fig.~2D, for relatively weak interaction ($h_r=8k_BT$) the model predicts that the stall torque initially increases linearly with IMF, consistent with tight coupling, but begins to decrease with IMF beyond a threshold as deterministic slipping sets in. Increasing $h_r$ shifts this threshold to larger IMF, extending the linear regime. Because experimentally measured stall torques across the physiological IMF range (triangles in Fig.~2D) remain linear, the BFM must operate in the tight-coupling regime, implying a lower bound $h_r\gtrsim12k_BT$.

Direct measurements of the rotor–stator interaction are difficult. However, our model allows $h_r$ to be inferred from measurements of their relative motion. Recent experiments by Hosu et al.~\cite{hosu2025torque} reported that the ratio of rotor to stator rotational speeds near stall is $6.2\pm0.5$ (speed $\sim3$ Hz; $\xi_r\approx37$ pN·nm·s in tethered-cell assays~\cite{wu2024torque}). With the gear ratio $\alpha=5/26$, this corresponds to an engagement fraction $\gamma\approx0.84\pm0.07$. Using \eqref{gamma-deltah}, we estimate the rotor–stator interaction strength to be $h_r\approx15.4\pm0.5\,k_BT$ (Fig.~2B).

Taken together with existing experiments, our analysis indicates that the BFM operates in a strong-coupling regime with rare thermally driven slipping events that produce only a negligible reduction in the maximum torque. In this regime, the stall torque satisfies $\tau_{\rm stall}\approx\tau_{\rm max}$ and is largely insensitive to microscopic details, providing a natural explanation for why the CW and CCW modes exhibit the same stall torque~\cite{yuan2008,Lo2013}.

\section{The torque-speed curve in the strong coupling regime}

\begin{figure}
    \centering
    \includegraphics[width=1.0\linewidth]{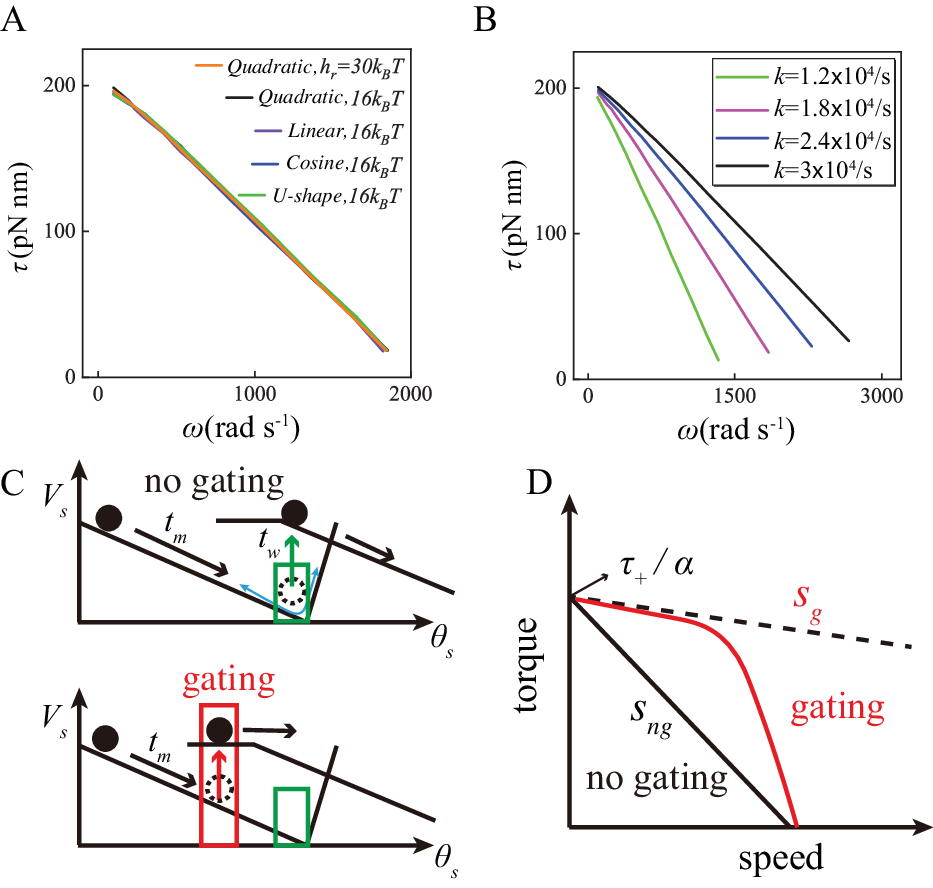}
        \caption{Torque-speed curves in the strong coupling regime. (A) The torque–speed curve is essentially unchanged over a broad range of rotor–stator interaction potentials $V_{r}$. (B) Varying the chemical transition rate $k$ shifts the speed but leaves the stall torque and the linear shape of the curve unchanged. (C) Stator mechanochemical dynamics. Top: In the absence of gating, after moving down the potential gradient generating torque, the stator dwells in a torque-free waiting phase at the bottom of the potential well (blue double arrow) before jumping to the next cycle (the green arrow). Results from panels A\&B are from this ``no-gating" model. Bottom: With gating, a transition (jumping) is triggered earlier along the downhill branch (red gate) bypassing the torque-free waiting phase near the minimum, which increases the torque-generating fraction of the cycle. $t_w$: waiting time; $t_m$: moving time. (D) Torque-speed curves with gating (red line) or no gating (black line). Gating suppresses the waiting phase and flattens the high-load slope, which produces a concave torque–speed curve.}
    \label{fig3}
\end{figure}

The above analysis indicates that the rotor and stator are strongly coupled under physiological conditions, leading to identical stall torques for CW and CCW rotation, consistent with experiments. However, the experimentally observed torque--speed relations differ in the intermediate-load regime. What gives rise to this CW--CCW asymmetry? Within our framework, two factors could in principle be responsible: the stator--rotor interaction potential $V_r$ or the ion-driven switching (jumping) rate $k(\theta_s)$.

We first examined the role of the stator--rotor interaction using the model shown in Fig.~\ref{fig1}C, in which ion translocation occurs only near the bottom of the stator potential (the ``no-gating'' model). As shown in Fig.~3A, the resulting torque--speed curve is largely insensitive to the specific form of $V_r$ provided the system remains in the strong-coupling regime with sufficiently large $h_r$. In this regime the torque--speed relation is linear, consistent with that observed in CW motors. Changing the jumping rate ($k$) simply shifts the maximum motor speed without altering the linear dependence (Fig.~3B). Thus, the no-gating model can account for the CW torque--speed relation but cannot explain the concave curve observed for CCW rotation.

To understand the origin of the linear torque--speed relation, we follow the approach of Meacci and Tu~\cite{Meacci2009} and partition each stator cycle into two phases: a \textit{moving phase}, during which the stator moves down its potential and generates torque $\tau_+$, and a \textit{waiting phase}, during which the stator fluctuates near the potential minimum without producing net torque. The waiting phase ends when ion-driven switching (``jumping'') to the next potential landscape occurs (Fig.~3C, top), initiating the next power stroke.

In the strong-coupling regime the average rotor torque satisfies
\begin{equation}
\langle\tau_r\rangle=\langle V_r'\rangle=\frac{\langle V_s'\rangle}{\alpha}-\frac{\xi_s}{\alpha^2}\langle\omega_r\rangle .
\label{tau_r}
\end{equation}
The duration of one cycle is $T=\Delta\theta_r/\langle\omega_r\rangle$, while the mean waiting time is set by the jumping rate $t_w\approx k^{-1}$. The stator therefore generates torque only during a fraction $(T-t_w)/T$ of the cycle, giving
\begin{equation}
\langle V_s'\rangle = \tau_+\frac{T-t_w}{T}
\approx \tau_+ - \frac{\tau_+ \langle\omega_r\rangle}{k\Delta\theta_r}.
\end{equation}
Substituting this expression in \eqref{tau_r} yields the linear torque--speed relation shown in Fig.~3D (black line):
\begin{equation}
\langle\tau_r\rangle=\tau_{\max}-s_{ng}\langle\omega_r\rangle,
\label{linear-tor-speed}
\end{equation}
with slope $
s_{ng}=\frac{\tau_+}{\Delta\theta_r \alpha k}+\frac{\xi_s}{\alpha^2}$ .

Experiments show that CCW motors generate higher torque than CW motors at intermediate loads, resulting in a flatter torque--speed curve near stall. Because the torque produced during the moving phase ($\tau_+$) is the same for both directions, this difference must arise from the waiting phase: CCW motors must spend less time in the torque-free waiting state. In the limiting case where the waiting time becomes negligible ($t_w\ll T$) at high load, $\langle V_s'\rangle\approx\tau_+$ and
\begin{equation}
\langle\tau_r\rangle
= \frac{\langle V_s'\rangle}{\alpha}-\frac{\xi_s}{\alpha^2}\langle\omega_r\rangle
\approx \tau_{\max}-s_g\langle\omega_r\rangle,
\label{upper-limit}
\end{equation}
with a much smaller slope $
s_g=\frac{\xi_s}{\alpha^2}\ll s_{ng}$ consistent with higher torque generation in CCW motors. 

What may shorten the waiting phase in CCW motors? One possible way is through a chemical “gating” mechanism~\cite{tu2018design}. In this scenario, a gating region (red region in Fig.~3C, bottom) located on the downhill part of the stator potential allows ion-triggered switching to the next cycle before the stator reaches the ``waiting" region near the potential minimum (green region in Fig.~3C, bottom) where it does not generate any net torque. By enabling early switching, the gate suppresses the waiting phase and maintains a higher average torque at high loads.

At low loads, however, the stator rotational speed is fast and the stator can pass the gating region quickly without switching (jumping). As a result, the effect of gating diminishes, the torque drops steeply with increasing speed, and the torque–speed curve converges to approximately the same maximum-speed point determined by the transition rate at the potential minimum as in the no-gating case. This mechanism naturally produces a concave torque–speed relation as illustrated in Fig.~3D (the red line).

Our analysis thus suggests an appealing explanation for the CW–CCW asymmetry: an additional gating mechanism may assist ion translocation in CCW motors, while being much weaker in CW motors. However, such a mechanism is difficult to reconcile with a single-step chemical transition driven by only one ion. In particular, if the stator switches in the middle of the potential ramp, it is unclear how further torque could be generated before the cycle completes, since no additional torque-generating pathway would be available. Remarkably, these difficulties are naturally resolved by recent structural results, which reveal that two ions alternately drive the stator. This picture provides a coherent mechanism for gated transitions while maintaining torque generation, as we discuss in the next section.


\section{A contact-dependent gating mechanism for ion release }

\begin{figure*}
    \centering
    \includegraphics[width=1\linewidth]{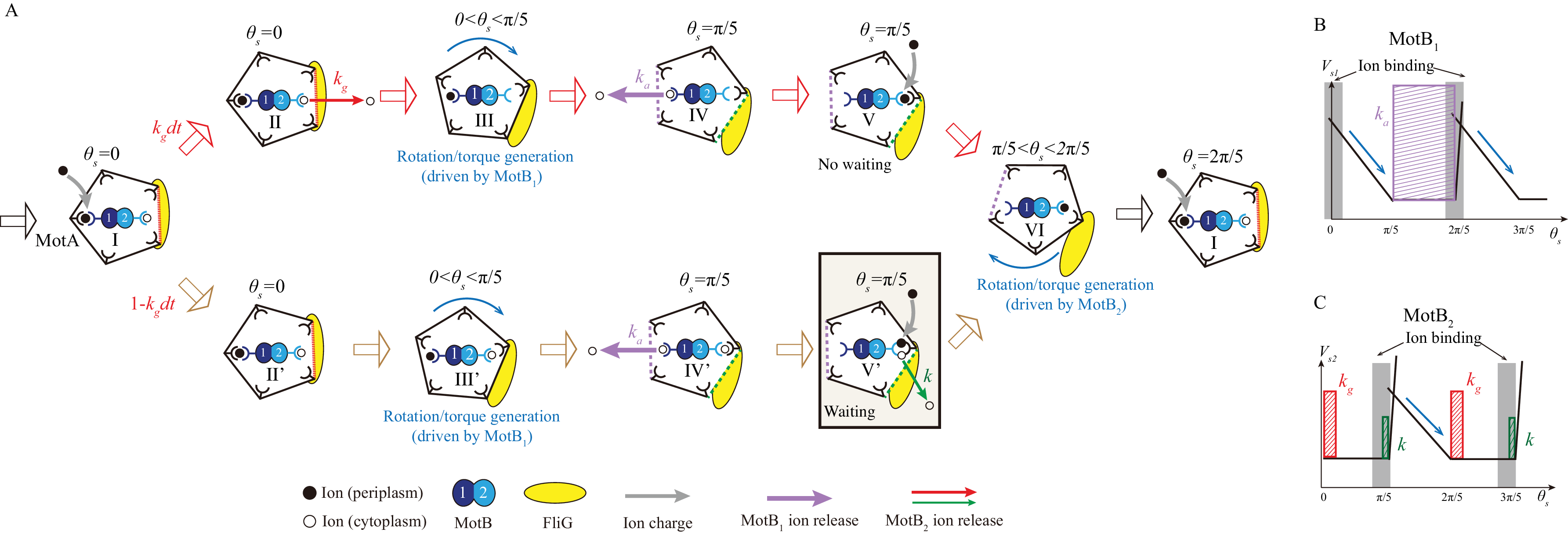}
    \caption{Mechanochemical dynamics of a stator unit during one torque-generating cycle ($\Delta \theta_s=2\pi/5$).  (A) Schematic of the two-ion pathways through the two MotB subunits. The branching of the two pathways (upper and lower) is determined by early release of ion at MotB$_2$ (proximal to FliG, light blue) with rate $k_g$. At $\theta_s=0$ (state I), ion from the periplasm binds to MotB$_1$ (distal from FliG, dark blue) when the MotA-MotB charging channel (half circle markers) is aligned (state I). If the ion on MotB$_2$  is released through the gating mechanism (red arrow), the cycle follows the upper branch (state II); otherwise it follows the lower branch (state II'). MotB$_1$ then drives a $\pi/5$  power stroke of the MotA ring (blue arrows; states III and III').  After the power stroke, the ion is released through the MotA-MotB release channel (dashed purple line, state IV and IV'). At $\theta_s=\pi/5$, an ion binds to MotB$_2$ and drives the next $\pi/5$ power stroke (state VI). If MotB$_2$ has not undergone early release, the system remains in the lower branch and waits until ion release occurs through a baseline pathway (rate $k$, state V'), generating a torque-free waiting phase. After a period ($2\pi/5$), the stator disengages from the current FliG and engages to the next FliG (not shown).  (B) Stator potential and ion kinetics for the distal MotB$_1$. Ion binds at $\theta_s=0$ and releases during $\pi/5<\theta_s<2\pi/5$ with rate $k_a$ (purple bar). (C) State potential and ion kinetics for the proximal MotB$_2$. Ion could be released either through the gating  channel with rate $k_g$ (red bar) at $\theta_s=0$ or near $\theta_s=\pi/5$ at the base-line rate $k$ (green bar).}
    \label{fig4}
\end{figure*}

We first describe a detailed model for torque generation in a stator unit consisting of a MotB dimer surrounded by a MotA pentamer, based on recent cryo-EM structures of the stator~\cite{santiveri2020,deme2020}. As shown in Fig.~4A, each MotB subunit forms an ion channel together with neighboring MotA subunits. Ion translocations through the two channels occur alternately and together drive rotation of the flagellar motor.

Each ion translocation cycle consists of two steps—ion binding and ion release—associated with two conformational states of MotB. When a MotB channel aligns with the channel of its neighboring MotA subunit ($\theta_s=0$ in Fig.~4A), an ion from the periplasm binds to MotB in its periplasm-facing conformation (P-state). Ion binding triggers a conformational switch to the cytoplasm-facing state (C-state), which raises the stator potential ($V_s$) to its peak value (“jumping”), powered by the ion motive force. During the subsequent power-stroke phase, relaxation of this interaction potential down the gradient of $V_s$ generates torque and drives a half-period rotation of the MotA ring ($\theta_s:0\rightarrow\pi/5$).

After the power stroke, the stator potential reaches its minimum and remains flat during the subsequent half-period ($\theta_s\in[\pi/5,2\pi/5]$). During this phase the bound ion must be released from the MotB C-state to the cytoplasm before the next translocation cycle can begin. Importantly, while one MotB subunit is in this flat phase, the other MotB subunit can bind an ion at $\theta_s=\pi/5$ and initiate its own power stroke, thereby sustaining continuous torque generation over a full cycle.

This alternating, out-of-phase operation of the two MotB channels is therefore crucial for continuous torque generation. Here we assume that ion binding is extremely rapid when a MotB channel aligns with its corresponding MotA channel (gray bars in Fig.~4B,C), so that the rate-limiting step is ion release. 
The release rate depends on the local MotA--MotB environment near the conserved acidic residue. We further assume this local environment can be affected by MotA--FliG contacts at the stator--rotor interface, allowing rotor engagement to gate ion release in a rotation-state-dependent manner. In this way, the mechanical contact between MotA and FliG is coupled to the chemical timing of ion release, providing a possible origin of different gating strengths in the CCW and CW states.

Consequently, the ion release rates of the two MotB subunits can differ significantly. For convenience, we assign the distal MotB$_1$, which is located away from the FliG ring, to the rapidly releasing pathway with rate $k_a$ in the flat regime ($\pi/5<\theta_s<2\pi/5$) following the power stroke (Fig.~4B). In contrast, we assign the proximal MotB$_2$, which faces the stator–rotor interface, to the contact-dependent pathway, with reduced release rate $k_g$. For simplicity, we place the release channel of MotB$_2$ near $\theta_s=0$ and $2\pi/5$, allowing a possible gated early release before the end of the flat regime ($\theta_s=2\pi/5$, Fig.~4C). We also include a baseline release pathway with rate $k$ at the end of the flat regime ($\theta_s=\pi/5,3\pi/5$) to prevent the system from becoming trapped. Note that this assignment is a convenient modeling choice rather than a unique structural requirement (the same mechanism would also apply if the contact-gated pathway were assigned to the distal MotB subunit instead). The essential assumption is that one release pathway is weakly affected by MotA--FliG contact and releases ions rapidly, whereas the other pathway is contact-gated.

This contact-dependent gating mechanism ($k_g$) for the proximal MotB$_2$ channel creates two distinct ion-translocation pathways.
In the upper pathway (Fig.~4A), the bound ion is released early, between $\theta_s=0$ and $\theta_s=\pi/5$, via gating. As a result, a new ion can bind immediately at the start of the next cycle ($\theta_s=\pi/5$), enabling continuous torque generation without delay. For simplicity, we assume that gating occurs immediately after torque generation by MotB$_2$ is completed at $\theta_s=0$. In contrast, if gating fails, the system follows the lower pathway. The ion remains bound until the end of the flat regime ($\theta_s=\pi/5$), blocking the binding of a new ion. This leads to a torque-free waiting phase, represented by the $V'$ state (shaded in Fig.~4A) near $\theta_s=\pi/5$. No torque is generated during this interval, which persists until the ion is eventually released through the slower baseline pathway at rate $k$.

The strength of the gating mechanism therefore determines how the cycle is partitioned between these two pathways. For larger values of $k_g$ (strong gating), the stator spends most of the cycle in torque-generating states, suppressing the waiting phase and driving the motor toward the upper-limit behavior described in \eqref{upper-limit}, which produces a concave torque–speed relation. In contrast, when $k_g$ is small (weak gating), early release rarely occurs and the system frequently dwells in the torque-free waiting state (state $V'$ in Fig.~4A), leading to the linear torque–speed relation described in \eqref{linear-tor-speed}.

\section{Differential contact-dependent gating explains the CW-CCW asymmetry}

\begin{figure}
    \centering
    \includegraphics[width=1.0\linewidth]{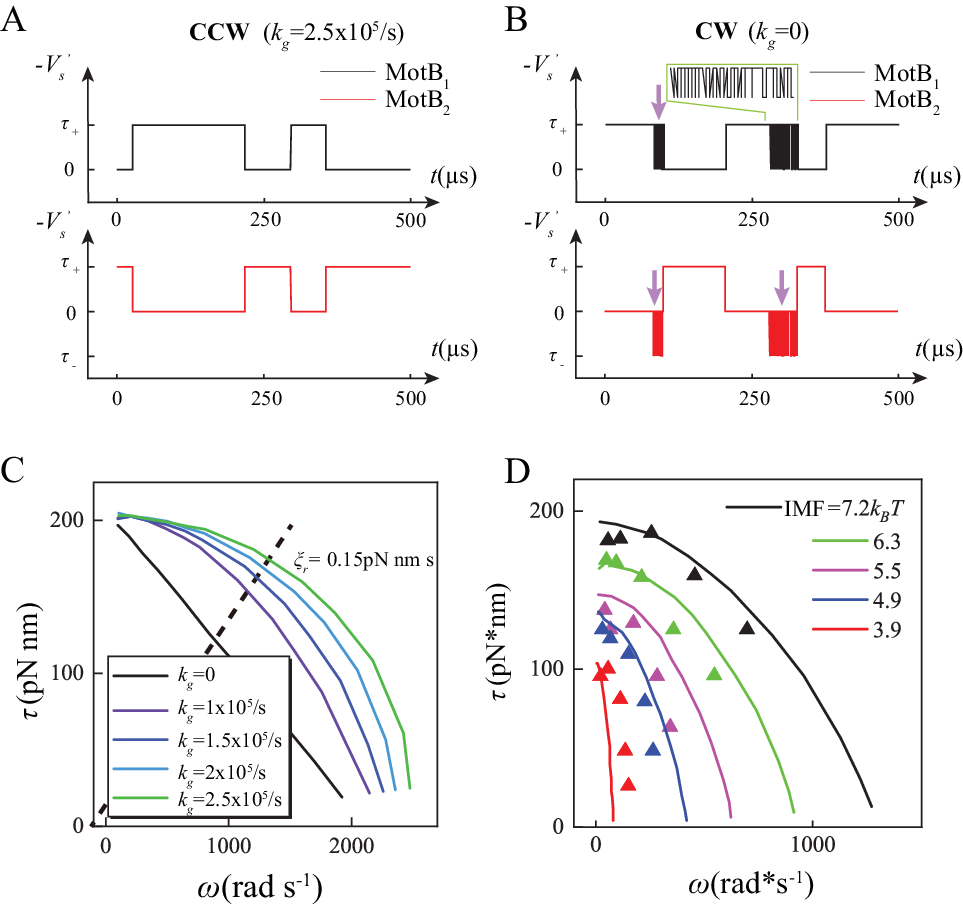}
\caption{Dynamics of instantaneous torque and the torque-speed relations for CCW and CW motors. The only difference for CCW and CW motors is: $k_{g,\text{CW}}=0$ and $k_{g,\text{CCW}}=k_g$. Other parameters used in simulations are given in SI.  (A) Instantaneous torque dynamics in CCW motors at high load ($\xi_r=0.15$ pN nm s). Gating keeps the stator predominantly in the positive-torque region. Ions on MotB$_1$ (black line) and MotB$_2$ (red line) driving the stator alternatively (switching between $\tau_+$ and 0). (B) Instantaneous torque dynamics in CW motors at high load. Near the equilibrium position (purple arrows, the waiting phase), MotB$_1$ alternates between $\tau_+$ and 0 (zoomed inset), and MotB$_2$ switches between at 0 and $\tau_-$, yielding zero net torque on average (state V' in Fig. 4). (C) Torque–speed relations with the ion translocation kinetics given in Fig.~4. The torque-speed curve is linear for CW motors with $k_g=0$. Increasing the gating rate $k_g$ enhances concavity of the torque–speed relation. (D) Torque-speed curves for CCW motors with different IMF agree with experimental data from \cite{Lo2013}. Here, we assume $\tau_+$ scales linearly with IMF as in Fig.~2D, and other parameters (e.g., $k_g$ and $k$) are given in the SI.}
\end{figure}

Our model immediately suggests that the distinct torque–speed curves observed for CCW and CW motors arise from different strengths of the gating mechanism, with $k_{g,\text{CCW}}\gg k_{g,\text{CW}}$. To test this hypothesis quantitatively, we performed numerical simulations of the model. Figures~5A and 5B show representative instantaneous torque traces for the two directions at high load, using $k_{g,\text{CW}}=0$ and $k_{g,\text{CCW}}=k_g\gg k$.

During CCW rotation, strong gating keeps the stator predominantly in the positive-torque region. MotB$_1$ and MotB$2$ alternately drive the motor, producing torque that switches between $\tau_+$ and 0 (Fig.~5A). In contrast, during CW rotation, where effective gating is absent (or weak), the stator spends a significant fraction of time fluctuating near its equilibrium position. In this regime MotB$_1$ generates positive or zero torque, while MotB$_2$ produces zero or negative torque. The resulting “tug-of-war’’ between the two MotB subunits traps the stator in a waiting state (Fig.~5B, purple arrows) with nearly zero net torque.

These microscopic move–wait dynamics give rise to the macroscopic torque–speed relations shown in Fig.~5C: the curve is concave when $k_g$ is large and becomes linear when $k_g$ is small, with the concavity increasing as $k_g$ increases. An analytical expression for the torque–speed relation is provided in the SI. We also examined the dependence of the torque–speed curve on the ion motive force (IMF) by scaling the stator potential $V_s$ with IMF. The resulting curves show excellent agreement with experimental measurements~\cite{Lo2013} (Fig.~5D and Fig.~S3–S4).

\section{Structural basis for the allosteric ion gating mechanism}

What is the mechanistic origin of the different gating strengths in CCW and CW motors? Cryo-EM studies have revealed pronounced structural differences between the two rotational states, most notably at the MotA–FliG contact interface~\cite{chang2020,tan2024,johnson2024,singh2024}. We hypothesize that these state-dependent contacts are transmitted allosterically to the MotA--MotB ion-release region, thereby modulating the ion-release gating rate. In this picture, the more favorable CCW interface would promote a larger effective gating rate than the CW interface, i.e., $k_{g,\text{CCW}}>k_{g,\text{CW}}$. Because the complete MotA--FliG contact geometry and the local environment near the MotB acidic residues are not fully resolved in static structures~\cite{tan2024,johnson2024,singh2024,chen2026}, we constructed CCW and CW interface models based on recently available high-resolution cryo-EM structures in \textit{Salmonella}. The MotAB stator architecture is conserved with the {\it E. coli} system, making these structures suitable templates for probing the MotA--FliG/MotA--MotB coupling mechanism\cite{Sowa2008}.  We then performed molecular dynamics (MD) simulations of the resulting models (Fig. 6 A\& B). We also considered different protonation states of MotB Asp25 (D25, corresponding to the conserved proton-binding homolog D32 in {\it E. coli} MotB) to test the robustness of local conformational response \cite{chen2026,luo2026}.

\begin{figure*}
    \centering
    \includegraphics[width=0.9\linewidth]{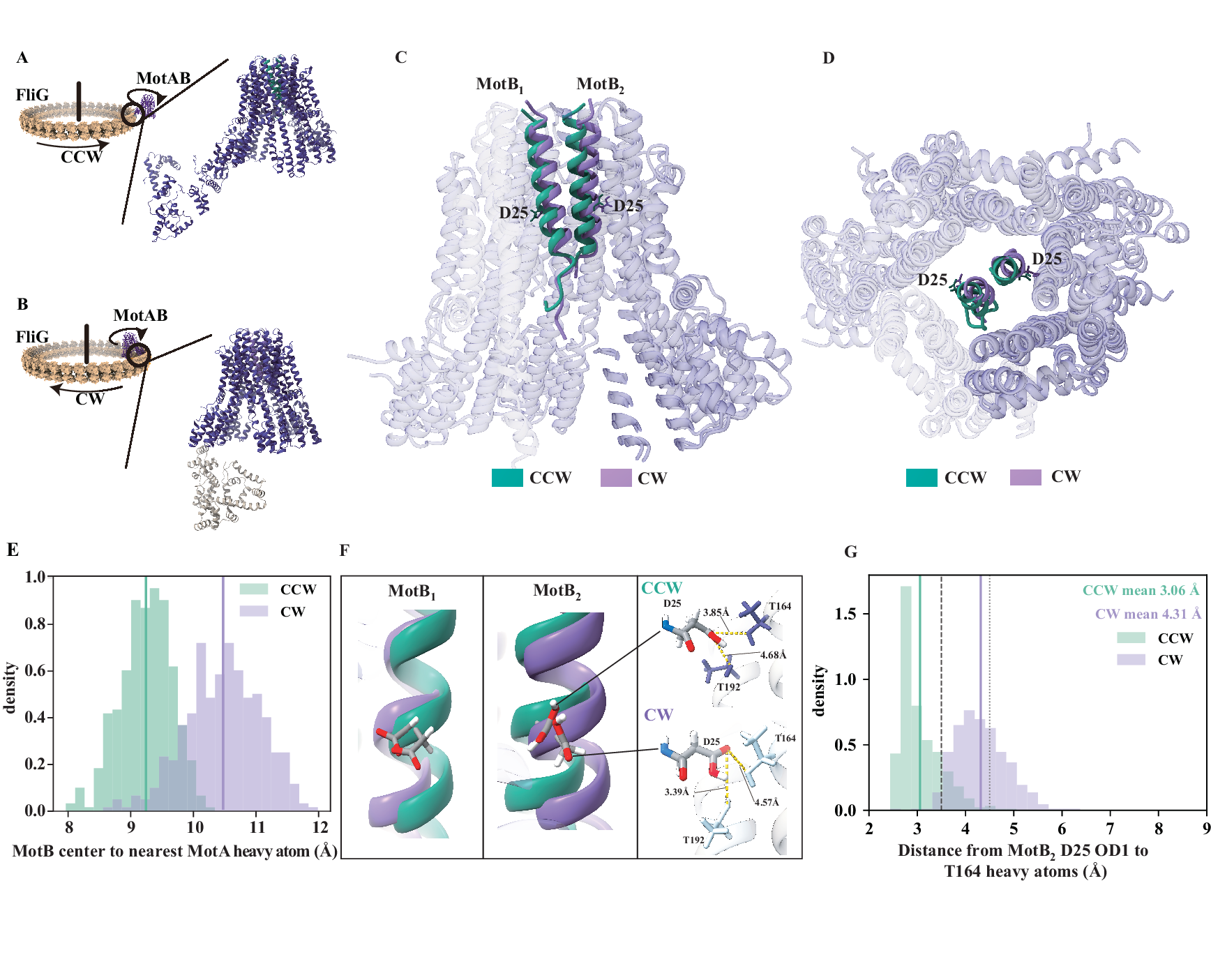}
    \caption{State-dependent MotA--FliG contacts are coupled to rearrangements of the MotA--MotB acidic-site geometry. (A,B) Structural models used for MD simulations of the MotA--FliG interface in the CCW (A) and CW (B) states, constructed from recent \textit{Salmonella} cryo-EM structures. The relative orientation of FliG and MotAB is indicated.
(C,D) Representative side (C) and top (D) views of the MotA--MotB acidic-site region in the stator complex, MotA subunits are aligned between CCW and CW states. The two MotB D25 residues are highlighted. Green: CCW; purple: CW.
(E) Distribution of the distance between the MotB acidic-site center and nearby MotA residues in CCW and CW simulations. The CCW state shows a smaller average distance, with a mean CCW--CW shift of -1.23~\text{\AA}.
(F) Local conformations of protonated D25 in the two MotB helices. The D25 side chain on MotB$_2$ (farther from FliG) exhibits a stronger state-dependent reorientation than the corresponding D25 on MotB$_1$ (closet to FliG). Representative CCW and CW conformations show different local contacts between MotB$_2$ D25 and nearby MotA residues, including T164 and T192.
(G) Distribution of the distance from protonated MotB$_2$ D25 OD1 to nearby MotA T164 heavy atoms. The shorter mean distance in CCW (3.06~\text{\AA}) than in CW (4.31~\text{\AA}) indicates a more hydrogen-bond-compatible D25--T164 geometry in the CCW state. 
}
    \label{fig5}
\end{figure*}

Our simulations reveal a state-dependent rearrangement of the MotA--MotB acidic-site region. In the CCW state, the acidic--site centers of the two MotB helices are closer to nearby MotA residues than in the CW state, with an average shift of 1.23~\text{\AA} (Fig. 6C, D \& E). This closer contact suggests that the CCW geometry strengthens local electrostatic and hydrogen-bond coupling between the MotB acidic residues and the MotA channel environment. Such coupling can favor proton release by shortening donor--acceptor distances and stabilizing hydrogen--bonded proton--transfer pathways, both of which are known to affect proton-transfer barriers and conduction rates~\cite{pomes1996,cukierman2000,freier2011,decoursey2024}. 

At the residual level, protonated D25 on one MotB helix farther from FliG in the initial configuration, which we assign as MotB$_2$, shows a pronounced state--dependent orientational change, whereas D25 on the other helix closest to the rotor interface, assigned as MotB$_1$, changes much less (Fig.~6F). This orientational change is accompanied by a shorter distance between MotB$_2$ D25 and MotA Thr164 (T164) in the CCW state than in the CW state, with mean distances of 3.06\AA\, and 4.31\AA, respectively (Fig. 6G). The shorter CCW distance places D25 in a more hydrogen-bond-compatible geometry with the MotA channel environment, which could facilitate local proton--transfer or release steps (see SI) \cite{chen2026,luo2026}.

Overall, our MD simulations suggest that FliG--MotA contact can allosterically modulate the MotA--MotB acidic-site region, with the strongest local response occurring at the MotB$_2$ D25 site in our structural assignment. The simulations do not directly measure the ion--release rate, but they identify two structural features expected to affect release kinetics: closer MotA--MotB coupling at the acidic-site region and a state--dependent D25 orientation that changes its local hydrogen--bond environment. These features are consistent with stronger ion--release gating in the CCW state than in the CW state, $k_{g,\text{CCW}}>k_{g,\text{CW}}$.


\section{Summary and Discussion}

Guided by extensive experimental data—including cryo--EM structures of the bacterial flagellar motor (BFM), direct observations of stator rotation~\cite{hosu2025torque}, and the distinct torque--speed relations measured for CW and CCW rotation, we developed a theoretical framework to uncover the mechanochemical principles underlying torque generation in the BFM. Our approach combines stochastic modeling of motor dynamics with MD simulations of key protein--protein interfaces to connect molecular structure with motor function.

We first developed a coarse-grained stochastic model describing the gear-like rotation between stator and rotor inspired by recent cryo-EM structures of the stator complex. The model shows that the macroscopic torque--speed relation is governed by two factors: the stator–rotor interaction potential and the kinetics of ion translocation through the stator. Comparison with experiments across different ion motive force (IMF) conditions and recent measurements of stator rotation indicates that the motor operates in a tight rotor–stator engagement regime under physiological conditions. In this regime, slipping between rotor and stator occurs only through rare thermal fluctuations. Such tight coupling ensures high energetic efficiency, as little of the IMF is dissipated through slippage.

A key insight from our analysis is that, within this strong-coupling regime, the torque–speed relation is largely insensitive to the detailed form of the rotor–stator interaction potential. Instead, the overall shape of the torque–speed curve is determined primarily by the kinetics of ion translocation. Guided by cryo-EM structures of the motor complex, we propose a contact-dependent ion-release mechanism in which MotA--FliG interactions modulate the local MotA--MotB environment near the MotB acidic site. Our MD simulations support this allosteric picture: the CCW and CW interfaces produce distinct acidic-site geometries, with the CCW state bringing the MotB acidic--site region closer to MotA and placing protonated MotB D25 in a more hydrogen--bond--compatible configuration. These structural differences suggest that ion release can be gated differently in the two rotational states. Our stochastic model shows that this difference in ion release kinetics naturally explains the distinct torque–speed relations observed experimentally: a linear torque–speed curve for CW motors and a concave torque–speed curve for CCW motors.

More broadly, our work highlights several mechano-chemical design principles underlying efficient torque generation in the BFM. First, tight rotor–stator coupling prevents loss of IMF through mechanical slippage. This design principle likely extends beyond the rotor–stator interface to other mechanical linkages in the motor, such as the FliG–FliM/N and rotor–hook connections. Indeed, cryo-EM structures show that the concentric rings of the rotor (including the LP and MS rings) form a highly rigid assembly~\cite{tan2021}. Second, efficient torque generation critically depends on the timing of ion translocation events. In particular, a gating mechanism that enables early ion release can suppress torque-free waiting phases and increase average torque output. At first sight such early release might appear problematic, since no torque would be generated between ion release and the next binding event. Remarkably, this challenge is resolved by the architecture of the stator itself: the MotB dimer forms two ion channels that operate alternately and out of phase during each stator rotation cycle ($2\pi/5$). This alternating operation enables continuous torque generation and naturally explains why stronger gating leads to higher torque output in CCW motors compared with CW motors.

Our model also leads to several experimentally testable predictions. First, mutations that weaken the electrostatic contacts at the MotA-–FliG torque transmission interface should reduce the effective coupling depth $h_r$, leading to lower stall torque and reduced motor efficiency~\cite{Blair1991,braun1999}. For example, residues such as MotA D86 (Chain~A3), E94 (Chain~A4), and FliG D241/D284/D289 form key charged contacts at the stator--rotor interface; mutations at these sites are expected to reduce $h_r$ and increase slippage under high load or high IMF, resulting in a measurable decrease in stall torque. Second, perturbations that alter the MotA--FliG interface or the MotA--MotB ion release pathway should change the effectiveness of early ion release and thereby modify the concavity (or knee speed) of the CCW torque–speed curve. For example, MotA T164 lies near MotB D25 in our structural model and may help stabilize a release-compatible local geometry. Mutating this residue is therefore predicted to perturb ion--release kinetics, producing defects in ion transmission and altering the CCW torque--speed relation (see SI). Third, comparison across different stator systems provide another test of the coupling picture. For example, the MotAB system in \textit{Pseudomonas aeruginosa} shows slipping at high load\cite{wu2024torque}, in contrast to the MotCD and \textit{E. coli} MotAB. In our framework, such behavior corresponds to weaker rotor--stator coupling ($h_r$), and direct rotor-stator tracking using the single--molecule approach developed in \cite{hosu2025torque} should reveal a lower engagement fraction ($\gamma$) at high load for MotAB compared to MotCD or \textit{E. coli} MotAB.

Finally, our study suggests several directions for future theoretical and computational work. Our current MD simulations reveal structural asymmetry at the MotA--FliG interface but do not yet resolve the detailed ion translocation process or the corresponding kinetic rates. It would therefore be valuable to extend these simulations to quantify ion release kinetics using explicit free-energy and rate calculations (e.g., enhanced-sampling MD or QM/MM approaches). Such studies could determine which physical factors at the interface—steric constraints, electrostatics, or allosteric coupling—dominate the gating effect. In addition, our present model treats stator components as rigid bodies, whereas MotA and MotB are flexible proteins that likely undergo conformational fluctuations. Incorporating such flexibility into coarse-grained models may be important for understanding additional motor properties such as load adaptation~\cite{wu2024torque,lele:2013:pnas,tipping:2013:mbio,nord:2017:catchbond,wadhwa:2019:pnas,wadhwa:2021:pnas}. Finally, this work focuses on the dynamics of a single stator unit. Extending the framework to motors containing multiple stators would allow the study of collective stator dynamics and rotational switching in the full motor assembly~\cite{wang2022direct,wu2024torque,yuan2008,Mattingly&Tu2026}.

\section{Acknowledgments}
J. Z., Y. H. and Y. C. are supported by the National Key Research and Development Program of China (Grant No.2024YFA0919600).
The Flatiron Institute is a division of the Simons Foundation.

\bibliography{motor_ref.bib}
\end{document}